# AI Explainability for Power Electronics: From a Lipschitz Continuity Perspective

Xinze Li, *Member, IEEE*, Fanfan Lin, *Member*, *IEEE*, Homer Alan Mantooth, *Fellow*, *IEEE*, and Juan J. Rodríguez-Andina, *Fellow, IEEE*.

***Abstract*—Lifecycle management of power converters continues to thrive with emerging artificial intelligence (AI) solutions, yet AI mathematical explainability remains unexplored in power electronics (PE) community. The lack of theoretical rigor challenges adoption in mission-critical applications. Therefore, this letter proposes a generic framework to evaluate mathematical explainability, highlighting inference stability and training convergence from a Lipschitz continuity perspective. Inference stability governs consistent outputs under input perturbations, essential for robust real-time control and fault diagnosis. Training convergence guarantees stable learning dynamics, facilitating accurate modeling in PE contexts. Additionally, a Lipschitz-aware learning rate selection strategy is introduced to accelerate convergence while mitigating overshoots and oscillations. The feasibility of the proposed Lipschitz-oriented framework is demonstrated by validating the mathematical explainability of a state-of-the-art physics-in-architecture neural network, and substantiated through empirical case studies on dual-active-bridge converters. This letter serves as a clarion call for the PE community to embrace mathematical explainability, heralding a transformative era of trustworthy and explainable AI solutions that potentially redefine the future of power electronics.**

## I. Introduction

Artificial Intelligence (AI) is increasingly permeating the lifecycle of power converters, credited to its capability in identifying complex patterns from data, automating decision-making, adapting to evolving environments, etc. Whereas black-box AI solutions are prevalent in power electronics (PE), their lack of explainability poses significant challenges to trust and adoption [1]. Explainable AI (XAI) has thus served as a cornerstone to confidently deploy AI in PE domains, supporting diversified applications like renewable energy integration [2], transportation electrification [3], smart cities, etc.

Emerging as the next generation of AI for engineering, physics-informed machine learning steers the AI learning process through integrating physical principles to safeguard AI explainability [4]. A noteworthy advancement in physics-informed machine learning for PE is the physics-in-architecture neural network (PANN), featuring a physically inspired recurrent neural architecture crafted from discretized state-space equations [5], providing physical explainability with enriched PE circuit insights whilst being light and flexible in nature [6], [7].

Despite its promise in offering PE-specific physical explainability, the mathematical explainability of PANN remains unexplored in the current literature. More broadly, the PE community lacks the awareness of prioritizing the mathematical foundation of AI-based solutions, impeding their widespread adoption in mission-critical applications, including but not limited to more electric aircrafts, submarines, and healthcare appliances. Consequently, this letter strives to establish standards and tools for evaluating the mathematical explainability of AI in PE, demonstrated through the validation of PANN's model stability and convergence under rigorous mathematical settings.

Overall, the mathematical explainability of PANN is justified from two critical aspects: inference stability and training convergence. Firstly, inference stability evaluates the consistency of model outputs in response to inputs and neural parameters, preventing abrupt reactions like exploding outputs to minor input variations. In real-time control and fault diagnosis of power converters, inference instability can lead to oscillations, overshoots, or false alarms caused by minor sensor fluctuations, compromising energy conversion reliability [8]. Secondly, training convergence ensures stable learning dynamics, such as a smoothly decreasing loss function during gradient descent. Divergent training can lead to suboptimal control or inaccurate diagnosis. In system-level applications like dynamic energy scheduling, a lack of training convergence may result in unrealistic dispatch, risking grid integrity [9]. These two metrics are universal indicators for assessing the mathematical explainability of data-driven AI models.

Centering on the proof of mathematical explainability, Lipschitz continuity is instrumental in validating both inference stability and training convergence. This property ensures that output or loss variations of data-driven AI models are proportionally bounded by input or neural parameter changes, with the associated Lipschitz constant quantifying the maximal gradient value, which effectively functions as gradient clipping to enforce training convergence.

To address the need for a rigorous theoretical foundation in AI models for the PE community, this letter emphasizes mathematical explainability from the perspective of Lipschitz continuity, and establishes a generic and comprehensive evaluation framework for mathematical explainability, including:

- **Inference stability:** Demonstrating the Lipschitz continuity of model outputs concerning inputs to ensure inference stability.
- **Training convergence:** Validating the Lipschitz continuity of loss functions with respect to neural parameters to justify training convergence.
- **Lipschitz-aware learning rates:** Proposing a strategy for selecting learning rates based on Lipschitz constants to accelerate convergence while mitigating overshoots and oscillations.

This letter calls upon the PE community to prioritize mathematical rigor in the development and evaluation of AI solutions, fostering trust and reliability in their deployment across PE applications.

## II. Mathematical Explainability of AI Models in Power Electronics and Lipschitz Continuity

### A. Definitions of Lipschitz Continuity and Lipschitz Constant

**Definition 1**. A function $f$: $\mathbb{R}^n \rightarrow \mathbb{R}^m$ is Lipschitz continuous if there exists a constant $L_1$ (the Lipschitz constant) such that [10]:

$$\|f(\boldsymbol{x_1}) - f(\boldsymbol{x_2})\| \le L_1 \|\boldsymbol{x_1} - \boldsymbol{x_2}\|, \forall \boldsymbol{x_1}, \boldsymbol{x_2} \in \mathbb{R}^n. \quad (1)$$

where $\|\cdot\|$ denotes a consistent norm, and $L_1$ quantifies the maximum rate of variation of $f$ w.r.t. $\boldsymbol{x}$ and serves as an upper bound for its magnitude. Lipschitz continuity in (1) is consistently used to prove inference stability and training convergence.

#### a) Generic Proof of Inference Stability

**Theorem 1**. An AI model satisfies inference stability if the gradient of model outputs $f(\mathbf{z})$ w.r.t. model inputs $\mathbf{z}$, namely its Jacobian matrix $\nabla_z f(\boldsymbol{z})$, is bounded, as formulated in (2) [11].

$$\|\nabla_z f(\boldsymbol{z})\| \le L_{1z} \xrightarrow{yield} \|f(\boldsymbol{z_1}) - f(\boldsymbol{z_2})\| \le L_{1z}\|\boldsymbol{z_1} - \boldsymbol{z_2}\| \quad (2)$$

$L_{1z}$ is the Lipschitz constant that captures the upper bound of the Jacobian matrix norm $\|\nabla_z f(\boldsymbol{z})\|$, safeguarding against over-amplified model responses to input variations for smoother behaviors. Inference stability enhances the robustness of AI models to input perturbations, which is vital in noisy environments and under risks of adversarial attacks. For instance, in real-world power converter control, sensor signals like output voltages are subjected to ambient noise and fluctuations [8], which could cause sharp changes in the duty cycle, leading to voltage oscillations, overshoots, and even control instability for an AI-based online controller without inference stability, jeopardizing mission success.

#### b) Generic Proof of Training Convergence

**Definition 2**. Training convergence is established [11] if the limit of the averaged regret approaches zero as the number of training epochs $T \rightarrow \infty$, as given in (3), where the regret is defined in (4).

$$\lim_{T\to\infty} \frac{Regret(T)}{T} = \lim_{T\to\infty} O\left(\frac{1}{\sqrt{T}}\right) = 0 \quad (3)$$

$$Regret(T) = \sum_{t=1}^{T} [f_t(\boldsymbol{\theta}_t) - f_t(\boldsymbol{\theta}^*)],$$

$$\text{where } \boldsymbol{\theta}^* = argmin_{\boldsymbol{\theta}\in\boldsymbol{\Theta}} \textstyle\sum_{t=1}^{T} f_t(\boldsymbol{\theta}). \quad (4)$$

The training convergence proof depends on the choice of optimizers. Here, the Adam optimizer is considered for analysis.

**Theorem 2**. Convergence of AI models trained with the Adam optimizer is achieved if the conditions in (5) hold [12], [13]:

$$\|\nabla_{\boldsymbol{\theta}} f_t(\boldsymbol{\theta})\|_2 \le G, \|\nabla_{\boldsymbol{\theta}} f_t(\boldsymbol{\theta})\|_\infty \le G_\infty$$

$$\|\boldsymbol{\theta}_n - \boldsymbol{\theta}_m\|_2 \le D, \|\boldsymbol{\theta}_n - \boldsymbol{\theta}_m\|_\infty \le D_\infty, \quad (5)$$

where, $G$ and $G_\infty$ are upper bounds on the 2-norm and infinity-norm of the gradient matrix of the loss function w.r.t. neural parameters $\boldsymbol{\theta}$. $D$ and $D_\infty$ bound the parameter updates. Small bounds can mitigate divergence, overshoots, and oscillations during training with a restricted search space.

The conditions in (5) imply that the gradient norm $\|\nabla_{\boldsymbol{\theta}} f_t(\boldsymbol{\theta})\|$ is bounded across all epochs, equivalent to the proof of Lipschitz continuity of loss functions, as indicated in (6).

$$\|\nabla_{\boldsymbol{\theta}} f(\boldsymbol{\theta})\| \le L_{1\boldsymbol{\theta}} \xrightarrow{yield} \|f(\boldsymbol{\theta_1}) - f(\boldsymbol{\theta_2})\| \le L_{1\boldsymbol{\theta}}\|\boldsymbol{\theta_1} - \boldsymbol{\theta_2}\| \quad (6)$$

Training convergence guarantees that AI models can accurately recognize informative and generalizable patterns from data. For example, for a reinforcement learning algorithm that is trained in real time, training convergence is crucial for reliable control performance for constantly varying conditions [14]. Similarly, in an AI-driven thermal management case, stable training ensures accurate thermal predictions [15], facilitating precise cooling and energy-efficient operation.

### B. Fundamentals of PANN: Formulation and Structure

Without loss of generality, this letter focuses on the proof of AI mathematical explainability in the context of the latest PANN models, which provide theoretical foundations complementary to PANN's physical explainability.

PANN, proposed by Li et al. in 2024 [5], integrates the general large-signal form of circuit state-space equations of power converters in (7) into its recurrent neural architecture through the discretization of numerical methods. In (7), $\boldsymbol{x}(t)$ and $\boldsymbol{u}(t)$ are time-dependent state and input variables, respectively, with associated circuit parameter-dependent matrices denoted as $A$ and $B$. Notably, the trainable neural parameters of PANN correspond to the circuit parameters $\boldsymbol{\theta}$. The formulation of PANN with the implicit Euler algorithm is expressed as (8), which underpins the recurrent neural structure shown in Fig. 1, where $\Delta t$ is the time interval. Notations defined in (9) are used throughout this letter, and their shapes are indicated in (10), where $D_x$, $D_u$, and $D_\theta$ are the dimensions of states, inputs, and circuit parameters, respectively.

$$\frac{d\boldsymbol{x}(t)}{dt} = A(\boldsymbol{\theta})\boldsymbol{x}(t) + B(\boldsymbol{\theta})\boldsymbol{u}(t) \quad (7)$$

$$\boldsymbol{x}[t_{k+1}] = [(1 - A\Delta t)^{-1} \quad (1 - A\Delta t)^{-1}B\Delta t] \begin{bmatrix} \boldsymbol{x}[t_k] \\ \boldsymbol{u}[t_{k+1}] \end{bmatrix} \quad (8)$$

$$\widehat{\boldsymbol{x}} \triangleq \boldsymbol{x}[t_{k+1}], \boldsymbol{z} \triangleq \begin{bmatrix} \boldsymbol{x}[t_k] \\ \boldsymbol{u}[t_{k+1}] \end{bmatrix}, \widehat{\boldsymbol{x}} = W(\boldsymbol{\theta})\boldsymbol{z} \quad (9)$$

$$\widehat{\boldsymbol{x}} \in R^{D_x\times 1}, \boldsymbol{u} \in R^{D_u \times 1}, \boldsymbol{z} \in R^{(D_x+D_u)\times 1},$$
$$W \in R^{D_x \times (D_x+D_u)}, \boldsymbol{\theta} \in R^{D_\theta \times 1} \quad (10)$$

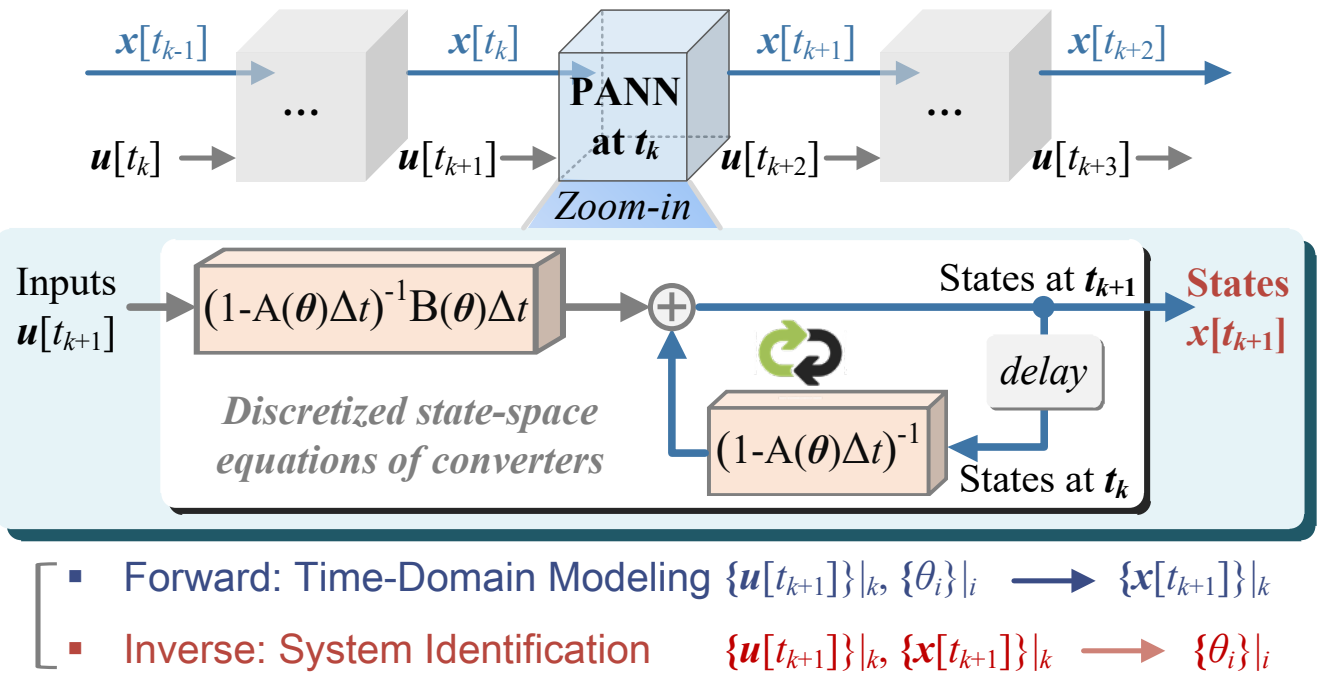

Fig. 1. The generic recurrent structure of PANN and its implications.

PANN models serve dual functionalities: 1) Forward execution for time-domain modeling of power converters in an unsupervised manner; 2) Inverse execution for system identification of unknown circuit parameters in a semi-supervised framework. PANN models have been successfully applied in a generative AI-empowered PE design paradigm (PE-GPT [16], proposed by Lin et al.), and advanced modulation optimization for the dual-active-bridge converter family [6], [7].

### C. Proof of Inference Stability of PANN

Building on Theorem 1, the inference stability of PANN models can be established by proving that the norm of the Jacobian matrix $\|\nabla_z \widehat{\boldsymbol{x}}(\boldsymbol{z})\|$ is bounded, as stated in (11). Lemmas 1.1 and 1.2 are utilized to support the proof of (11).

$$\|\nabla_z \widehat{\boldsymbol{x}}(\boldsymbol{z})\| = \|W(\boldsymbol{\theta})\| \le L_{1z} \quad (11)$$

**Lemma 1.1** For stable power converters, the norms of the state and input matrices in (7), $\|A\|$ and $\|B\|$, are bounded due to the physical constraints of circuit parameters and stability condition.

**Lemma 1.2** Given the boundedness of $\|A\|$ as stated in Lemma 1.1, there always exists a sufficiently small $\Delta t$ such that $\|A\Delta t\| < 1$. Using the Neumann series expansion [10], $\|(1\text{-}A\Delta t)^{-1}\|$ is bounded.

Lemmas 1.1 and 1.2 collectively contribute to the boundedness of the norms of both entries in the Jacobian matrix $W(\boldsymbol{\theta})$ in (8), thereby proving the inference stability of PANN models. With inference stability, PANN models achieve accurate time-domain modeling of stable power converters, serving as a mathematical proof of the experimentally validated waveforms in [5].

### *D. Proof of Training Convergence of PANN*

**Theorem 3.** If the conditions in Theorem 2 are satisfied, the regret over $T$ epochs, $Regret(T)$, for PANN models trained with Adam is bounded as expressed in (12) [12]. The average regret, $Regret(T)/T$, decreases at a rate of $\mathrm{O}(1/\sqrt{T})$, indicating that the accuracy of PANN improves with training epochs, although the rate of improvement diminishes as $T$ increases.

$$Regret(T) \le \frac{dD_\infty{}^2 G_\infty \sqrt{1-\beta_2}}{2\alpha(1-\beta_1)(1-\lambda)^2} + \frac{dD^2 G_\infty}{2\alpha(1-\beta_1)}\sqrt{T} + \frac{\alpha(1+\beta_1)dG_\infty{}^2}{(1-\beta_1)\sqrt{1-\beta_2}(1-\gamma)^2}\sqrt{T} = \mathrm{O}(\sqrt{T}) \quad (12)$$

Based on Theorem 3, the training of PANN models achieves convergence by justifying the boundedness of the gradient norm $\|\nabla_{\boldsymbol{\theta}} f_t(\boldsymbol{\theta})\|$ and the parameter updates $\|\boldsymbol{\theta}_n - \boldsymbol{\theta}_m\|$. Since PANN is designed to characterize the time-domain behaviors of power converters, the root mean square error (RMSE) defined in (13) is commonly used, where $x^*$ is the target states. The gradient of the RMSE loss w.r.t. neural parameters $\boldsymbol{\theta}$ is derived in (14). Using the Cauchy-Schwarz inequality [11], the gradient norm is shown to be bounded by (15).

$$f(\boldsymbol{\theta}) = 0.5(\hat{\boldsymbol{x}} - \boldsymbol{x}^*)^T(\hat{\boldsymbol{x}} - \boldsymbol{x}^*) \quad (13)$$

$$\nabla_{\boldsymbol{\theta}} f(\boldsymbol{\theta}) = \frac{\partial f}{\partial \hat{\boldsymbol{x}}}\left(\frac{\partial \hat{\boldsymbol{x}}}{\partial W}\frac{\partial W}{\partial \boldsymbol{\theta}}\right) = (\hat{\boldsymbol{x}} - \boldsymbol{x}^*)^T\left(\boldsymbol{z}\frac{\partial W}{\partial \boldsymbol{\theta}}\right) \quad (14)$$

$$\|\nabla_{\boldsymbol{\theta}} f(\boldsymbol{\theta})\| \le \|W - W^*\| \cdot \|\boldsymbol{z}\|^2 \cdot \left\|\frac{\partial W}{\partial \boldsymbol{\theta}}\right\| = L_{1\boldsymbol{\theta}} \quad (15)$$

Prior to proving the training convergence of PANN using Theorem 2, Lemma 2 and Assumption 1 are introduced.

**Lemma 2** The neural parameters $\boldsymbol{\theta}$ of PANN, which correspond to the circuit parameters of power converters, are physically bounded, implying that $\|\boldsymbol{\theta}_n - \boldsymbol{\theta}_m\|$ is bounded.

**Assumption 1** The norms of the derivative of the Jacobian matrix w.r.t. $\boldsymbol{\theta}$, $\|\partial W/\partial \boldsymbol{\theta}\|$, are bounded. This is a standard assumption for power converters with robust stability.

(11) provides the bounds for the first term in (15), and the second term $\|\mathbf{z}\|$ is also bounded due to its physical interpretation. Along with Lemma 2 and Assumption 1, the conditions in (5) are satisfied, proving the training convergence of PANN. Since the training of PANN corresponds to identifying unknown circuit parameters, its convergence implies an asymptotically stable system identification. Moreover, smaller Lipschitz constants $L_{1\boldsymbol{\theta}}$ in (15) result in smoother training dynamics, such that during the transient phase of identification, the estimates of circuit parameters $\boldsymbol{\theta}$ are free from oscillations and overshoots.

### *E. Lipschitz-Aware Selection of Learning Rates for PANN*

In pursuit of optimal training performance for PANN, a Lipschitz-aware strategy to select learning rates is proposed in (16), where $\alpha_i$ is the individual learning rate for the $i^{th}$ parameter $\theta_i$, with $\theta_{i,min}$ and $\theta_{i,max}$ being the lower and upper limits of $\theta_i$. The second-order Lipschitz constant, $L_{2\theta}$, which captures the upper bound of the Hessian matrix norm $\|\nabla^2_{\boldsymbol{\theta}} f(\boldsymbol{\theta})\|$, is defined in (17) and (18).

$$\alpha_i = \mathrm{O}\big(G_\infty \cdot (\theta_{i,max} - \theta_{i,min})/L_{2\boldsymbol{\theta}}\big) \quad (16)$$

$$\nabla_{\boldsymbol{\theta}}{}^2 f(\boldsymbol{\theta}) = \left(\boldsymbol{z}\frac{\partial W}{\partial \boldsymbol{\theta}}\right)^T\left(\boldsymbol{z}\frac{\partial W}{\partial \boldsymbol{\theta}}\right) + (\hat{\boldsymbol{x}} - \boldsymbol{x}^*)^T\left(\boldsymbol{z}\frac{\partial^2 W}{\partial \boldsymbol{\theta}\partial \boldsymbol{\theta}^T}\right) \quad (17)$$

$$\|\nabla^2_{\boldsymbol{\theta}} f(\boldsymbol{\theta})\| \le \|\boldsymbol{z}\|^2 \cdot \left\|\frac{\partial W}{\partial \boldsymbol{\theta}}\right\|^2 + \|W - W^*\| \cdot \|\boldsymbol{z}\|^2 \cdot \left\|\frac{\partial^2 W}{\partial \boldsymbol{\theta}\partial \boldsymbol{\theta}^T}\right\| = L_{2\theta} \quad (18)$$

This Lipschitz-aware heuristic strategy ensures stable and efficient training convergence for PANN, as highlighted below: Firstly, the second-order gradient is bounded by $L_{2\theta}$, which is incorporated into the denominator of $\alpha_i$, preventing oscillations in large-curvature regions and stabilizing training [17]. Secondly, the first-order Lipschitz constant $G_\infty$ in the numerator accounts for the gradient scale, enabling effective updates of $\boldsymbol{\theta}$. Thirdly, to cater for major differences in parameter magnitudes, $\alpha_i$ is scaled proportionally to the range of the $i^{th}$ parameter to explore the parameter space effectively.

## III. Case Study: Dual-Active-Bridge Converters

### *A. PANN Model for Dual-Active-Bridge (DAB) Converters*

This section analyzes the mathematical explainability of a PANN model for DAB converters, widely applied in solid state transformers [18], energy storage systems, etc. The state-space equation is given in (19), where $i_L$ is the key state variable, $v_p$ and $v_s$ are input variables, and the circuit parameters $\boldsymbol{\theta}_{dab}$ include $L_k$, $R_L$, and $n$. Using (8), the discretized state-space equation is derived in (20), and the recurrent PANN model for the DAB converter, customized in Fig. 2, is detailed in Table I.

$$\frac{di_L(t)}{dt} = -\frac{R_L}{L_k}i_L(t) + v_p(t) - nv_s(t) \quad (19)$$

$$\hat{\boldsymbol{x}}_{dab} = i_L[t_{k+1}] = \frac{[L_k \quad \Delta t \quad -n\Delta t]}{L_k + R_L\Delta t}\begin{bmatrix} i_L[t_k] \\ v_p[t_{k+1}] \\ v_s[t_{k+1}] \end{bmatrix} = W_{dab}(\boldsymbol{\theta}_{dab})\boldsymbol{z}_{dab} \quad (20)$$

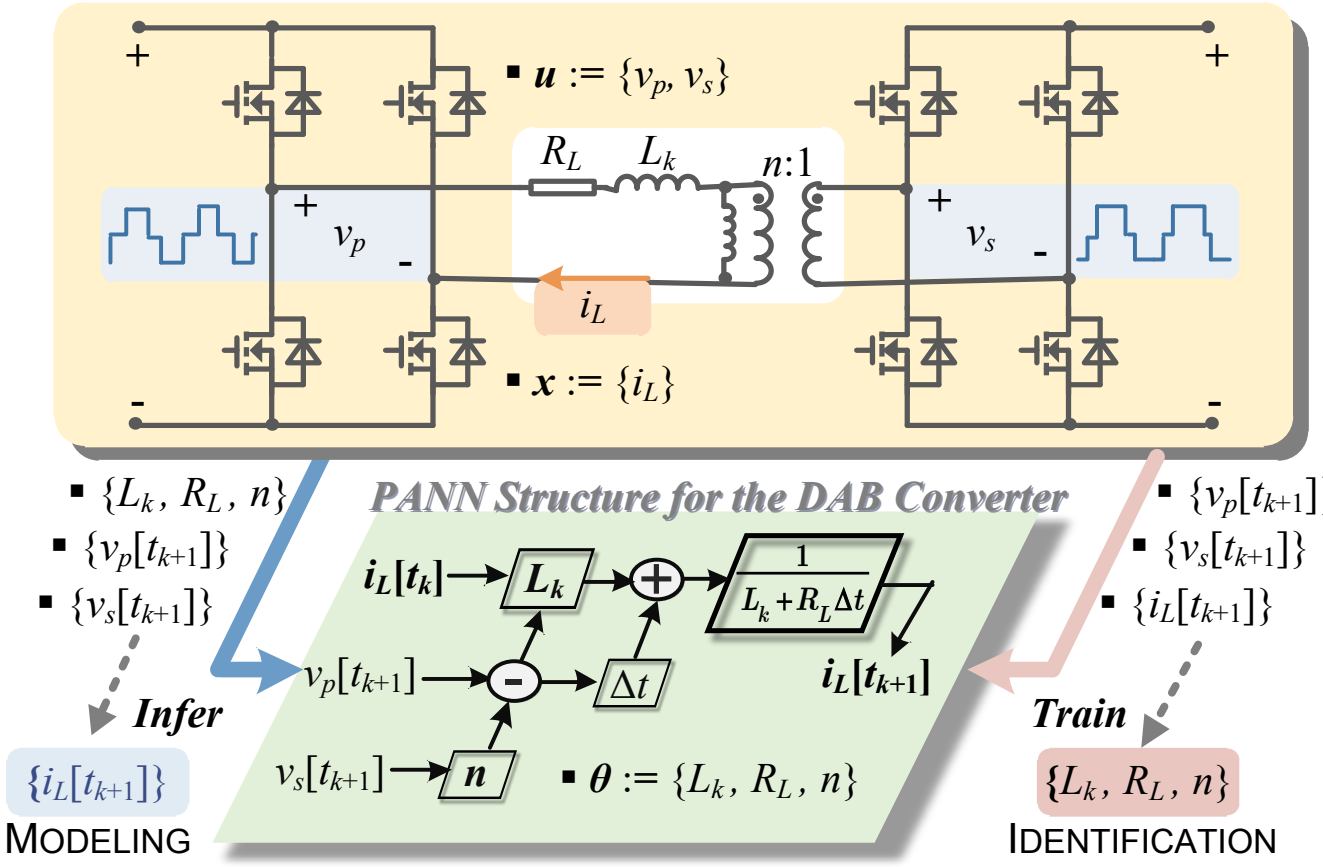

Fig. 2. DAB converters and its PANN model.

TABLE I. Configurations of the PANN for the DAB Converter

| | |
|---|---|
| Inputs: $\boldsymbol{z}_{dab} := \{i_L[t_k], v_p[t_{k+1}], v_s[t_{k+1}]\}$ | Outputs: $\hat{\boldsymbol{x}}_{dab} := i_L[t_{k+1}]$ |
| Parameters: $\boldsymbol{\theta}_{dab} := \{L_k, R_L, n\}$, $f_s$ = 50 kHz, $\Delta t$ = 80 ns | |
| Train-test-validation data size: 2, 50, 50 | Learning rates $\alpha_i \in [10^{-7}, 10^1]$ |
| Ground-true neural (circuit) parameters $\boldsymbol{\theta}^*_{dab}$: $L_k$ = 63 μH, $R_L$ = 1.8 Ω, $n$ = 1.0 | |
| Parameter ranges: $L_k \in$ [10 μH, 200 μH], $R_L \in$ [10 mΩ, 3 Ω], $n \in$ [0.8, 1.2] | |

### *B. Inference Stability of the PANN for DAB Converters*

As derived in (20), the boundedness of the Jacobian matrix norm $\left\|\nabla_{\boldsymbol{z}_{dab}}\hat{\boldsymbol{x}}_{dab}(\boldsymbol{z}_{dab})\right\|$, $W_{dab}$, is validated, with its upper

bound being the Lipschitz constant $L_{1z}$, which is close to but less than 1 under the infinity norm $\|\cdot\|_\infty$. The theoretical Lipschitz constant $L_{1z}$ is empirically validated via Monte Carlo (MC) simulations, as shown in Fig. 3.

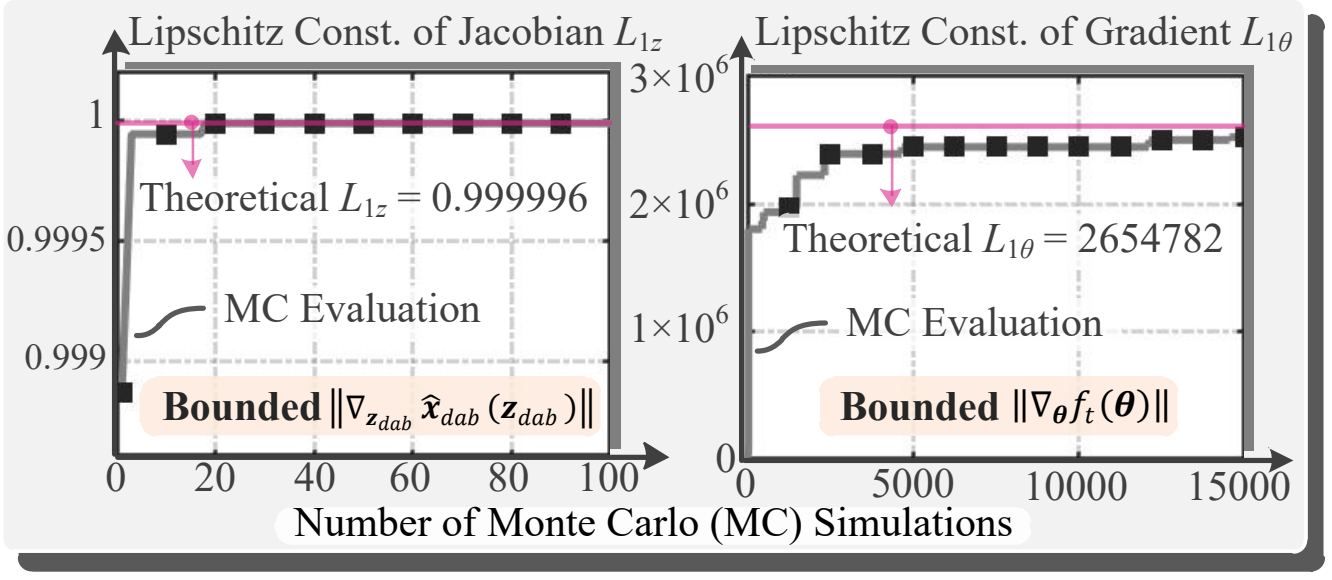

Fig. 3. Evaluation of Lipschitz constants through MC simulations.

### C. Training Convergence of the PANN for DAB Converters

According to Theorems 2 and 3, training convergence of the PANN model is ensured if the norms of the Jacobian matrix derivatives, $\|\partial W_{dab}/\partial \boldsymbol{\theta}_{dab}\|$, are bounded. As derived in (21), these matrix entries depend on the circuit parameters $\boldsymbol{\theta}_{dab}$, so physical constraints are naturally imposed to bound their norms. The upper bound of $\|\partial W_{dab}/\partial \boldsymbol{\theta}_{dab}\|$ is primarily influenced by the derivatives w.r.t. $L_k$, as indicated by the entries in the first column of (21).

$$\frac{\partial W_{dab}}{\partial \boldsymbol{\theta}_{dab}} = \frac{\Delta t}{(L_k + R_L \Delta t)^2} \begin{bmatrix} R_L & -L_k & 0 \\ -1 & -\Delta t & 0 \\ n & n\Delta t & -(L_k + R_L \Delta t) \end{bmatrix} \tag{21}$$

Similarly, MC simulations have been conducted to validate the theoretical Lipschitz constant of the RMSE loss w.r.t. neural parameters $\boldsymbol{\theta}$, $L_{1\theta}$ or $G_\infty$, as shown in Fig. 3. The empirically validated boundedness of $L_{1\theta}$ further justifies training convergence.

Besides, to illustrate training convergence, Fig. 4 presents the training performance for various learning rates. The training dynamics, including regrets, losses, and parameter estimates under Strategies 1, 3, and 5, are shown in Fig. 5. Low learning rates (Strategy 1) imply slow convergence, whereas high rates (Strategy 5) exhibit undesired overshoots and oscillations due to regions of high loss curvature. In comparison, the proposed Lipschitz-aware Strategy 3 is free of such concerns, whilst attaining the fastest convergence and high modeling accuracy. In addition, an ablation study (Strategy 6) demonstrates the effectiveness of scaling individual $\alpha_i$ to the parameter range, without which, convergence slows down, and overshoots and oscillations occur.

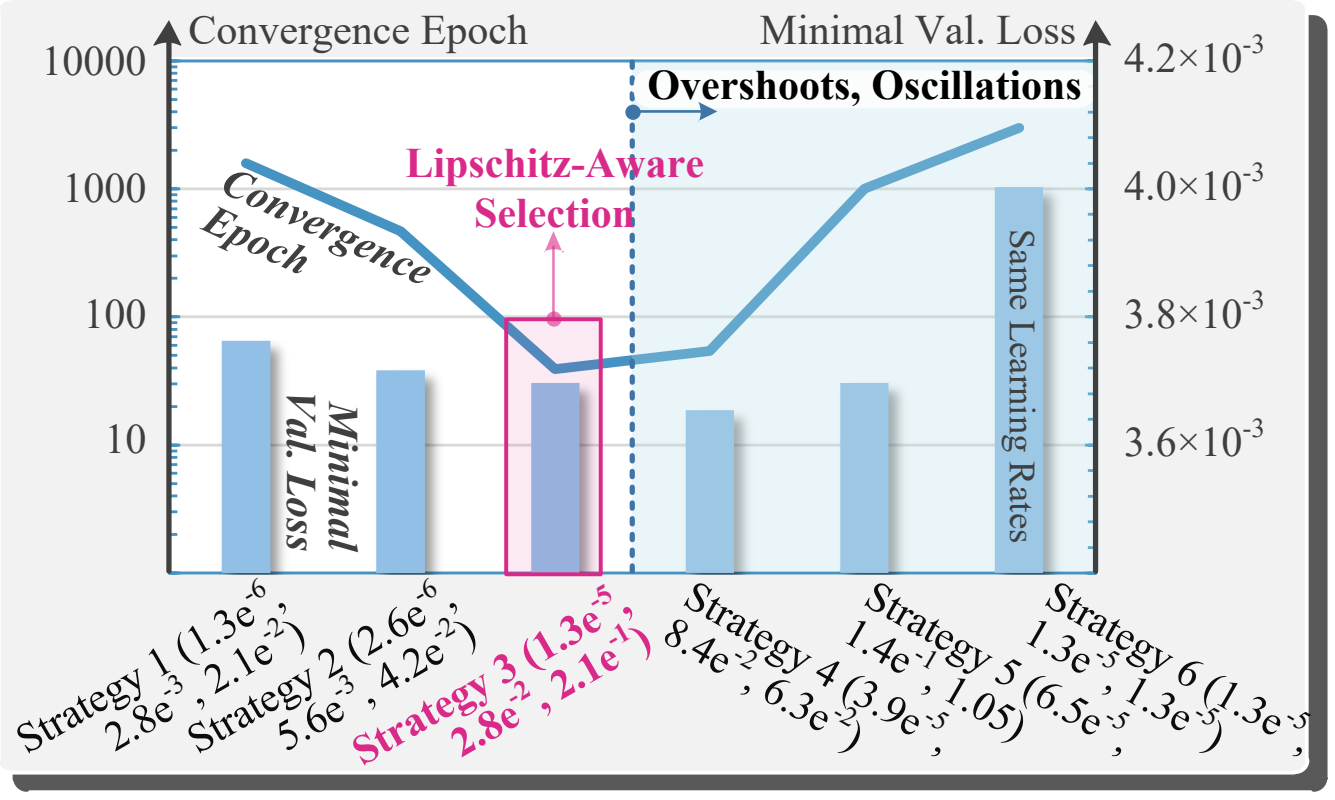

Fig. 4. Training performances for different learning rates $\alpha_i$. Values in (·) indicate $\alpha_i$ for $L_k$, $n$, and $R_L$. Learning rates in Strategy 3 are tuned based on Lipschitz constants $L_{1\theta}$ and $L_{2\theta}$.

Strategies 1, 3, and 5, representing low, optimal, and high learning rates, are analyzed. In Fig. 5 (a), nontrivial oscillations are observed in Strategy 5, whereas Strategy 1 shows the slowest loss reduction process. Estimates for the neural parameter $L_k$ during training are plotted in Fig. 5 (b), where Strategy 5 exhibits an 87% overshoot, and the Lipschitz-aware selected Strategy 3 converges in 39 epochs – the fastest among all. In Fig. 5 (c), the increase of $Regret(T)$ follows the square root of epochs $T$, consistent with (12), and the average regret approaches 0 as $T$ increases, confirming model convergence.

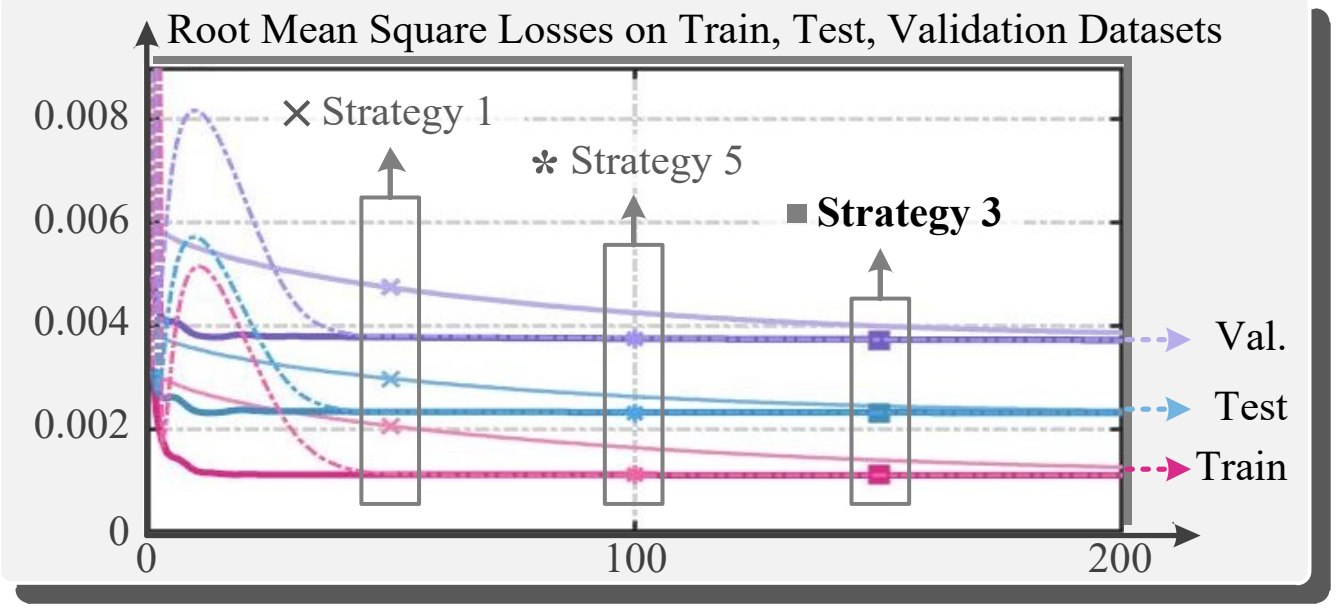

(a)

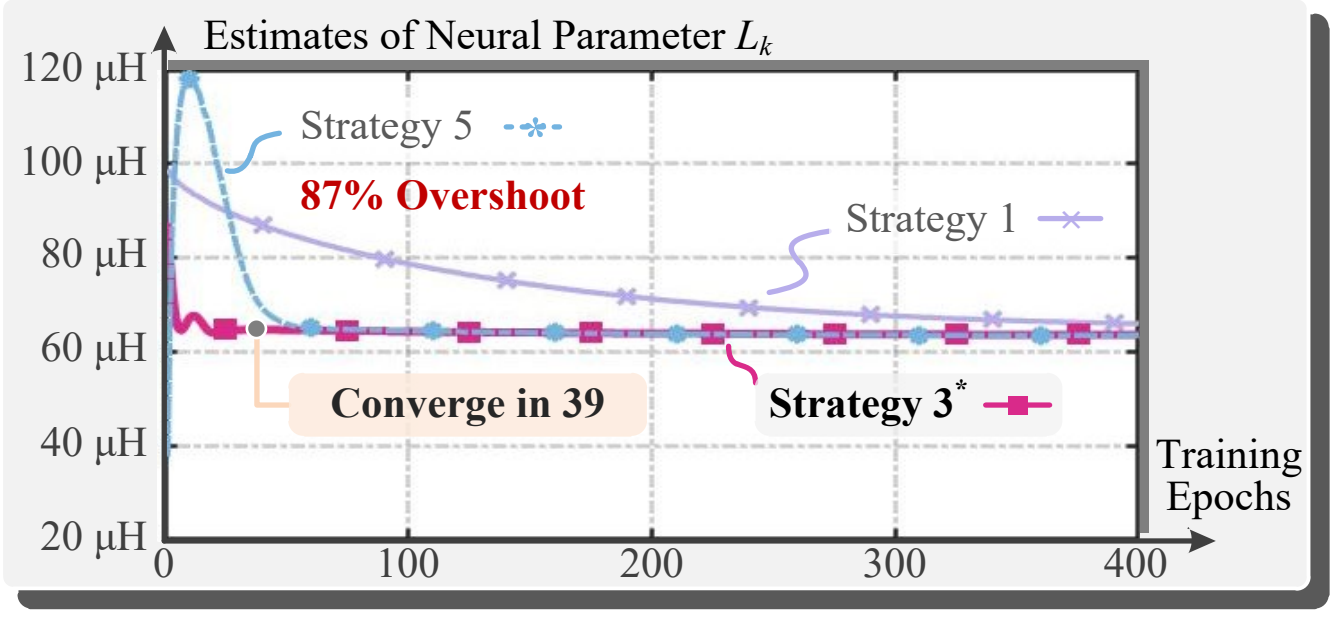

(b)

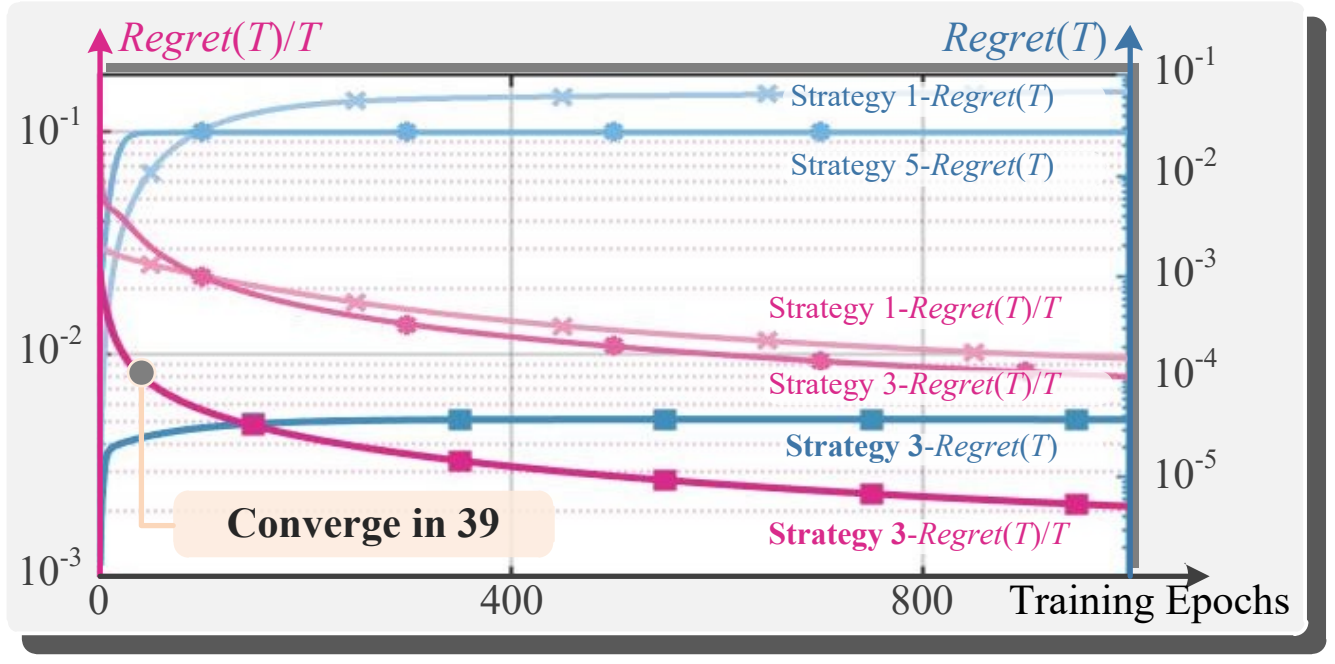

(c)

Fig. 5. Training dynamics of PANN under Strategies 1, 3, and 5: (a) RMSE losses; (b) Estimates for the neural parameter $L_k$; (c) Regret and average regret curves.

In a nutshell, the MC simulations empirically validate the inference stability and training convergence of the PANN for DAB converters. The bounded Lipschitz constants, $L_{1z}$ and $L_{1\theta}$, indicate the coherent and stable model output $\hat{\boldsymbol{x}}_{dab}$ and RMSE loss for neighboring inputs $\boldsymbol{z}_{dab}$ and parameters $\boldsymbol{\theta}_{dab}$. Furthermore, the observed training dynamics across various learning rates justify the time complexity of convergence with Adam, following $O(1/\sqrt{T})$, where the Lipschitz-aware strategy achieves the fastest convergence while preserving smoothness.

## IV. Conclusion

This letter is the first attempt to systematically address the mathematical explainability of AI models in power electronics, focusing on inference stability and training convergence from the perspective of Lipschitz continuity. The proposed generic framework emphasizes two aspects. First, the sufficient condition of inference stability is the Lipschitz continuity of model outputs with respect to inputs, validated through the boundedness of the Jacobian norm. Second, training convergence relies on the Lipschitz continuity of loss functions concerning neural parameters, verified via the boundedness of the gradient norm. Besides, a Lipschitz-aware strategy for learning rate selection is introduced to enhance convergence speed and mitigate oscillations and overshoots, consisting of stabilization with the second-order Lipschitz constant, effective gradient scaling, and adaptation to parameter magnitudes.

The framework's universality is demonstrated on a cutting-edge physics-in-architecture neural network (PANN), bridging a critical gap in the literature with rigorous theoretical proof of its mathematical explainability. Case studies on a PANN model for DAB converters empirically validate the framework's feasibility and the effectiveness of the Lipschitz-aware learning rate selection strategy.

This work aspires to establish a benchmark and raise awareness for mathematical explainability of AI-driven solutions in power electronics, fostering trust in mission-critical applications. By leveraging Lipschitz continuity as a cornerstone, we safeguard the reliability and robustness of AI, paving the way for their increasing penetration in power electronics.